\documentclass[aps, pra, reprint, superscriptaddress, twocolumn, showkeys, amsmath, amssymb, longbibliography]{revtex4-1}
\usepackage[english]{babel}
\usepackage{amsmath, amssymb, bbm, mathrsfs, bm, braket, color, graphicx, comment, amsfonts, dsfont}
\usepackage[colorlinks,citecolor=blue,urlcolor=blue]{hyperref}

\newcommand{\pd}{{\phantom{\dag}}}

\begin{document}
\title{Quantum phase transitions and a disorder-based filter in a Floquet system}
\author{Balaganchi A. Bhargava}
\affiliation{IFW Dresden and W{\"u}rzburg-Dresden Cluster of Excellence ct.qmat, Helmholtzstrasse 20, 01069 Dresden, Germany}
\author{Sanjib  Kumar  Das}
\affiliation{IFW Dresden and W{\"u}rzburg-Dresden Cluster of Excellence ct.qmat, Helmholtzstrasse 20, 01069 Dresden, Germany}
\author{Ion Cosma Fulga}
\affiliation{IFW Dresden and W{\"u}rzburg-Dresden Cluster of Excellence ct.qmat, Helmholtzstrasse 20, 01069 Dresden, Germany}

\begin{abstract}
Two-dimensional periodically-driven topological insulators have been shown to exhibit numerous topological phases, including ones which have no static analog, such as anomalous Floquet topological phases.
We study a two dimensional model of spinless fermions on a honeycomb lattice with periodic driving.
We show that this model exhibits a rich mixture of weak and strong topological phases, which we identify by computing their scattering matrix invariants.
Further, we do an in-depth analysis of these topological phases in the presence of spatial disorder and show the relative robustness of these phases against imperfections.
Making use of this robustness against spatial disorder, we propose a filter which allows the passage of only edge states, and which can be realized using existing experimental techniques.
\end{abstract}

\maketitle

\section{Introduction}

In recent years, a great amount of interest has been devoted to the field of topological insulators \cite{Thouless1982, Bernevig2006a, Bernevig2006b, Hasan2010, Qi2011}. 
Generally, topological insulators are defined as having a gapped bulk spectrum, but supporting lower dimensional gapless states, which can reside on surfaces or edges of the system. 
The existence of the gapless states is a consequence of the well known bulk-boundary correspondence principle, meaning the number of those boundary states will be dictated by a topological number, which is calculated from the bulk eigenstates of the system. 
This topological number, also known as a topological invariant, characterizes the nontrivial nature of the bulk, which is deeply connected to the symmetries of the system.
In this context, many earlier pioneering works have also classified the topological insulators based on the dimension and symmetries of the system, leading to the so called periodic table of topological phases \cite{Schnyder2008, Kitaev2009, Ryu2010, Chiu2016}.     

Recently, the understanding of topology has also been extended to periodically-driven systems, which feature rich topological phases that have no counterparts in static topological systems. 
The periodically-driven systems which realize topological phases are called Floquet topological insulators (FTI) \cite{Kitagawa2010, Lindner2011, Cayssol2013, Lindner2013, Rudner2013, Lababidi2014, Leykam2016, Bomantara2016, Yao2017, Oka2019}. 
The discovery of FTIs has enriched the earlier periodic table of topological insulators \cite{Else2016, Roy2017}. 
FTIs have also been realized in multiple experimental setups, such as in photonic waveguide lattices \cite{Rechtsman2013, Mukherjee2017, Maczewsky2017}, coupled ring resonators \cite{Guo2020}, and ultracold atoms \cite{Wintersperger2020}. 

One of the unique features of FTIs is the possibility of so-called anomalous topological phases, in which the bulk bands are topologically trivial, meaning that the system can host robust extended states even when \emph{all} states in the bulk of the system are localized. 
When the localization of bulk states is a consequence of disorder, these systems are called anomalous Floquet Anderson insulators (AFAI) \cite{Titum2016, Nathan2017, Kundu2020}. 
A study of some of the localization properties leading to the formation of AFAIs has been established using a periodically driven Kitaev model \cite{Kitaev2006, Fulga2019}. 
More recently, the generation of an AFAI phase in the presence of spatial disorder has been studied in a two-dimensional Chern insulator driven by onsite potential kicks \cite{Liu2020} or by an oscillating electric field \cite{Zhang2021}.
However, a detailed study of how the strength and type of spatial disorder impact the totality of the phase diagram of an FTI is missing. 
So, it is naturally of interest to understand the stability and robustness of FTI phases in the presence of disorder.  

In this article we address this question of robustness of FTI on a honeycomb lattice, a model which we define in section \ref{sec:model}.  
Then, in section \ref{sec:phases} we explore the phases present in the model. 
We calculate the invariants in each of the phases. 
In section \ref{sec:disorder}, a systematic  study of the robustness of the system to spatial disorder is presented. 
To achieve this, we provide a way to construct the Floquet operator in real space, analytically, even when the system is disordered.
Using the insight gained with respect to the relative robustness of the different topological phases, in section \ref{sec:filter} we propose a filter that can be realized in photonic crystals. 
This filter can be used to improve the signal corresponding to the propagating topological modes by removing the spurious contributions of the bulk states. 
We conclude in section \ref{sec:conc}.

\section{The Model}\label{sec:model}

\begin{figure}[t]
\includegraphics[width=0.68\columnwidth]{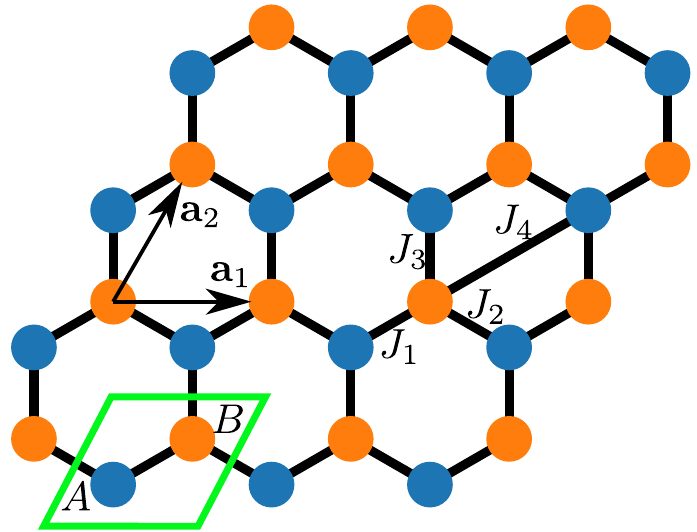}
\caption{
The honeycomb lattice. 
The unit cell is shown using a green parallelogram.
The blue circles and orange circles denote the sublattice $A$ and $B$, respectively. 
The four kinds of hopping are marked as $J_{1}$, $J_{2}$, $J_{3}$ and $J_{4}$, which are active in the time slices 1, 2, 3, and 4, respectively. 
To avoid clutter, only one of the $J_4$ hoppings is shown.
The Bravais vectors $\textbf{a}_1$ and $\textbf{a}_2$ are shown in one of the hexagons.
}
\label{fig:lattice_hop}
\end{figure}

We consider a system of spinless fermions on a honeycomb lattice with a time-dependent hopping amplitude. 
In addition to the nearest-neighbor hoppings, we consider a hopping across each hexagonal plaquette, see Fig.~\ref{fig:lattice_hop}. 
The Hamiltonian is given by
\begin{equation}\label{eq:main_ham}
{\cal H} = \mu\sum_{i, \alpha} c^{\dagger}_{i,\alpha}c^\pd_{i,\alpha}  +\sum_{i, j, \alpha, \beta} J_{(i,j)_{\alpha,\beta}}(t) (c^{\dagger}_{i,\alpha}c^\pd_{j,\beta} +  {\rm h.c.}),
\end{equation}
where $(i,j)_{\alpha,\beta}$ denotes a pair of sites connected by a hopping $J$, and $t$ is time.
The Latin indices denote the unit cell of the triangular lattice, while the Greek indices denote the sublattice degree of freedom, corresponding to two sites per unit cell.
$c^{\dagger}_{i,\alpha}$ and $c^\pd_{i,\alpha}$ are the usual fermionic creation and annihilation operators.
We consider four types of hopping throughout the lattice, $J_{1,2,3,4}$, as shown in Fig.~\ref{fig:lattice_hop}. 
The hopping amplitude $J(t)$ is a function of time and has the following form in one complete driving period, $T$:
\begin{equation}
\begin{split}
J(t) = 
\begin{cases}
J_{1} = J_{s}, & \quad J_{2}=J_{3}=J_{4}=0, \quad \\
& \text{for}\quad nT < t \leq nT + \frac{T}{4}, \\
\\
J_{2} = J_{s}, & \quad J_{1}=J_{3}=J_{4}=0, \quad \\
& \text{for}\quad nT + \frac{T}{4} < t \leq nT + \frac{T}{2}, \\
\\
J_{3} = J_{s}, & \quad J_{1}=J_{2}=J_{4}=0, \quad \\
& \text{for}\quad nT + \frac{T}{2} < t \leq nT + \frac{3T}{4}, \\
\\
J_{4} = J_{d}, & \quad J_{1}=J_{2}=J_{3}=0, \quad \\
& \text{for}\quad nT + \frac{3T}{4} < t \leq nT + T, \\
\end{cases}
\end{split}
\end{equation}
where $J_s$ and $J_d$ are constants, $n\in\mathbb{Z}$, and the onsite term is set to $\mu=0$. 

The spectrum of the above system in the case of a time-independent hopping can be obtained exactly, by using periodic boundary conditions along both directions.
We Fourier transform the Hamiltonian Eq.~\eqref{eq:main_ham} using the following definitions:
\begin{equation}\label{eq:k_ham}
\begin{split}
c_{i,\alpha} & =\frac{1}{\sqrt{2N}} \sum_{\textbf{k}} e^{-i\textbf{k}\cdot\textbf{r}_{i}}c_{\textbf{k}, \alpha} \\
& \text{and} \\
c^{\dagger}_{i,\alpha} & =\frac{1}{\sqrt{2N}} \sum_{\textbf{k}} e^{i\textbf{k}\cdot\textbf{r}_{i}}c^{\dagger}_{\textbf{k}, \alpha}.
\end{split}
\end{equation}
Here, $N$ denotes the total number of unit cells, ${\bf k}$ is the two-dimensional momentum vector, and $\textbf{r}_{i} = n_1 \textbf{a}_1 + n_2\textbf{a}_2$, with $n_{1,2}$ integers and $\textbf{a}_{1,2}$ the Bravais vectors (see Fig.~\ref{fig:lattice_hop}).
The momentum-space Hamiltonian is
\begin{gather}
{\cal H} = 
\sum_{\textbf{k}}
\begin{pmatrix}
 c^{\dagger}_{\textbf{k}, A} &  c^{\dagger}_{\textbf{k}, B}
\end{pmatrix}
H({\bf k})
\begin{pmatrix}
c^\pd_{\textbf{k}, A} \\  c^\pd_{\textbf{k}, B}
\end{pmatrix}, \\
H({\bf k}) = 
\begin{pmatrix}
0 &  \mathcal{J}({\bf k})\\ 
\mathcal{J}^{*}({\bf k}) & 0
\end{pmatrix},\label{eq:Hmom}
\end{gather}
where $\mathcal{J}({\bf k}) = J_{1} + J_{3}e^{ik_2} + J_{2}e^{ik_1} + J_{4}e^{i(k_1+k_2)}$, and $k_j = {\bf a}_j \cdot {\bf k}$.
The eigenvalues are given by $E = \pm |\mathcal{J}|$.

Now turning our attention to the driven system, we study the Floquet operator of the system and its properties.
The time evolution of a system from time $t_{1}$ to time $t_{2}$ is governed by the time-evolution operator given by
\begin{equation}
U(t) = \mathcal{T} e^{-i\int_{t_{1}}^{t_{2}} H(t) dt },
\end{equation}
where $H(t)$ is the time-dependent Hamiltonian, $\mathcal{T}$ denotes time ordering, and we have set $\hbar=1$. 
When the system evolves periodically in time, the time-evolution operator over one driving period is called the Floquet operator.
For the specific time drive that we have, the Floquet operator is a product of the time-evolution operators in each time slice:
\begin{equation}\label{eq:fq.def}
\begin{split}
F = \mathcal{T} e^{-i\int_{0}^{T} H(t) dt } & = e^{-iH_{4}\frac{T}{4}}e^{-iH_{3}\frac{T}{4}}e^{-iH_{2}\frac{T}{4}}e^{-iH_{1}\frac{T}{4}} \\
& = F_{4}F_{3}F_{2}F_{1}.
\end{split}
\end{equation}

The eigenvalue equation governed by the Floquet operator is
\begin{equation}\label{eq:floquet.def}
F \ket{\psi} = e^{-i\epsilon T} \ket{\psi} .
\end{equation}
Since the Floquet operator is unitary, the eigenvalues are phases. 
The $\epsilon$ in Eq.~\eqref{eq:floquet.def} is called the quasi-energy, and it is defined modulo $\frac{2\pi}{T}$. 

With the above form of the Floquet operator, when the quasi-energy is plotted as a function of momentum, we obtain the Floquet band structure. 
The Floquet bands are therefore periodic in both momentum and quasi-energy. 
The Floquet band structure is similar to the bands of static systems. 
In the system that we study, there are two quasi-energy gaps, at $\epsilon T= 0$ and $\epsilon T= \pi$. 

As long as $\mu = 0$, the system obeys particle-hole symmetry (PHS). 
This can be seen as follows. 
The Hamiltonian $H({\textbf{k})}$ in Eq.~\eqref{eq:Hmom} belongs to class BDI \cite{Altland1997} and is particle-hole symmetric:
\begin{equation}
\mathcal{P}^{-1}H(\textbf{k})\mathcal{P} = -H(-\textbf{k}),
\end{equation}
where $\mathcal{P} = \sigma_{z}\mathcal{K}$ is the PHS operator, $\sigma_z$ is the third Pauli matrix in sublattice space, and $\mathcal{K}$ denotes complex conjugation.
The manifestation of this in the Floquet operator can be worked out as 
\begin{equation}
\mathcal{P}^{-1}F(\textbf{k})\mathcal{P} = F(-\textbf{k}).
\end{equation}
As a result, in the Floquet band structure, one can find a state at $(-\epsilon, -{\bf k})$ corresponding to a state at $(\epsilon, {\bf k})$. 
 
 \begin{figure*}[tb]
\includegraphics[width=\textwidth]{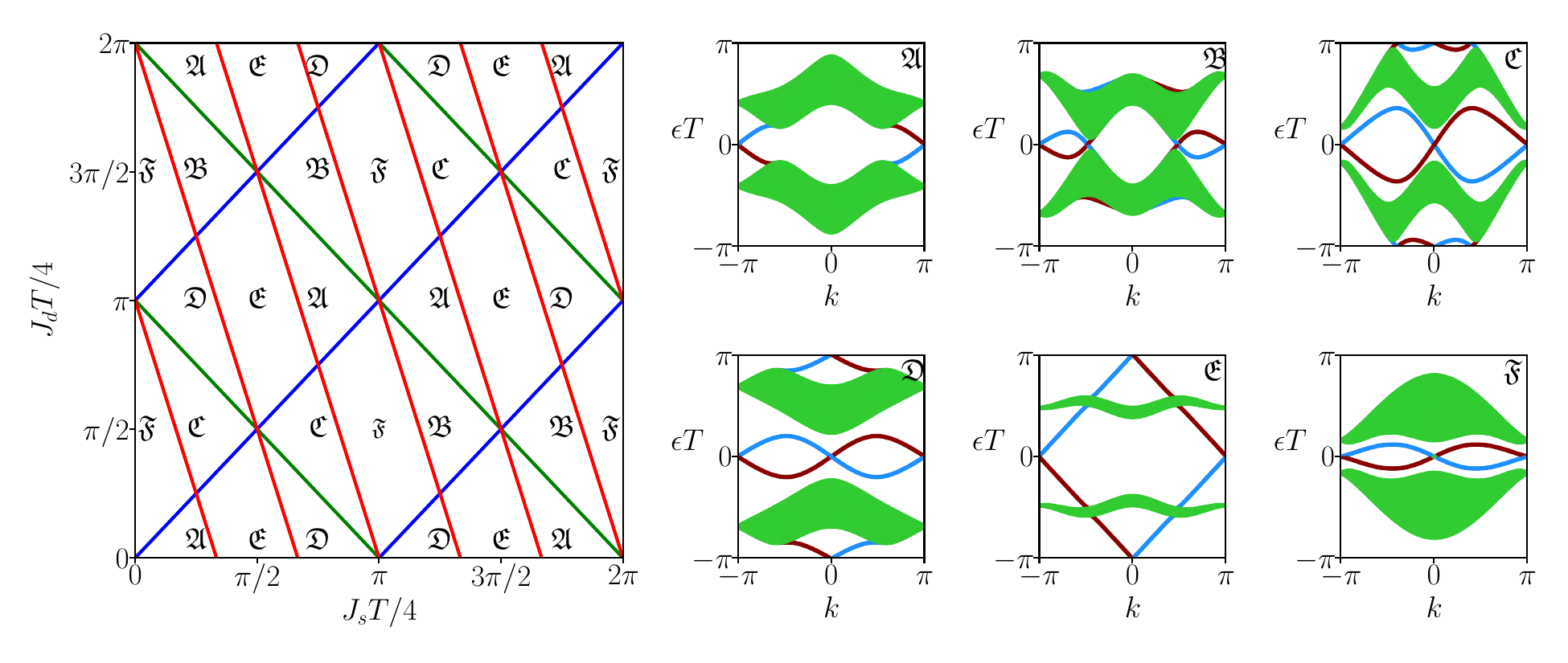}
\caption{
Left: The analytically obtained phase diagram for the Floquet system, showing the bulk gap closing lines as $J_{s}$ and $J_{d}$ are varied.
There are a total of six different topological phases, separated by families of straight lines, as indicated in Eq.~\eqref{eq:phase_lines}. 
The red lines correspond to $k_{1,2}=0$, the blue lines correspond to $k_{1,2}=\pi$ and the green lines correspond to $k_{1,2}=\mp\pi/2$ gap closing points in the Brillouin zone.
Right: Band structures for all the six phases in a ribbon geometry: infinite along ${\bf a}_1$, and having a width of 50 unit cells along ${\bf a}_2$. $k$ denotes the dimensionless momentum along the infinite direction of the ribbon. 
Each panel has has been labeled according to the phase that it corresponds to. 
Panels $\mathfrak{A}$, $\mathfrak{B}$, $\mathfrak{C}$, $\mathfrak{D}$, $\mathfrak{E}$, and $\mathfrak{F}$ are plotted at $(J_{s}T/4, J_{d}T/4)=(1, 6)$, $(0.85, 5)$, $(1, 1.55)$, $(0.85, 3)$, $(1.5, 3)$, and $(0.5, 1)$.
The colors in the band structures show where the states are localized along the $\textbf{a}_2$ direction. 
The green color corresponds to a state that is present in the bulk. 
The red color corresponds to a state on the bottom edge and blue corresponds to a state on the top edge.}
\label{fig:phase_diag}
\end{figure*}
 
\section{Phase Diagram}
\label{sec:phases}

By varying the values of $J_{s}$ and $J_{d}$, the Floquet system realizes a series of topological phases. Between them, topological phase transitions are signaled by the closing and reopening of either (or both) of the $\epsilon T = 0$ and $\epsilon T=\pi$ bulk gaps. We show these phases and phase transitions in Fig.~\ref{fig:phase_diag}. All our numerical results are obtained using the kwant code \cite{Groth2014}, and are available as supplemental material files at \cite{code}.

Due to PHS, the bulk gaps of the Floquet operator given by Eq.~\eqref{eq:fq.def} closes whenever $\epsilon T=n \pi$ for some ${\bf k}$, with $n\in \mathbb{Z}$. We find that these gap closings occur only at $k_{1,2}=0,\pm\pi/2,\pm\pi$, when
\begin{equation}
\begin{split}
(J_{d} + 3J_{s})\frac{T}{4} & = n\pi \quad \text{for} \quad (k_1, k_2)=(0,0), \\
(J_{d} - J_{s})\frac{T}{4} & = n\pi \quad \text{for} \quad (k_1, k_2)=(\pi,\pi), \\
(J_{d} + J_{s})\frac{T}{4} & = n\pi \quad \text{for} \quad (k_1, k_2)=(-\pi/2,\pi/2).
\end{split}
\label{eq:phase_lines}
\end{equation} 
Therefore, the gap closings are families of straight lines in the $J_s$ -- $J_d$ plane.
Also, note that the phase diagram is periodic in both the hopping variables $J_{s}$ and $J_{d}$:
\begin{equation}
F(J_{s}, J_{d}) = F(J_{s}, J_{d} + \frac{8\pi}{T}) = F(J_{s}+\frac{8\pi}{T}, J_{d}).
\end{equation}

Each of the phases in the phase diagram is characterized by a set of strong and weak topological indices. 
Since the characterization of the phases involves the study of edge states, we now shift to a ribbon geometry.
We consider a system which is infinite along the $\textbf{a}_1$ direction, but has a finite width along $\textbf{a}_2$. 

The strong index is a $\mathds{Z}$ invariant and measures the net number of chiral edges states on a given edge at a particular quasi-energy gap.
Since there are two quasi-energy gaps, there are two strong indices for each phase. 
We denote them $\mathcal{W}_{0}$ and $\mathcal{W}_{\pi}$, corresponding to the strong index at the $\epsilon T=0$ and $\epsilon T=\pi$ quasi-energy gaps, respectively.
These invariants are related to the Chern numbers of each of the two bulk bands by the relation \cite{Rudner2013}
\begin{equation}\label{eq:strong.invariant.relation}
C_{\pm} = \pm( \mathcal{W}_{\pi} - \mathcal{W}_{0}),
\end{equation}
where $C_{+}$ is the Chern number of the band at $\epsilon T>0$ in our convention, and $C_{-}$ is the Chern number of the band at $\epsilon T<0$.

The weak indices, on the other hand, are $\mathds{Z}_{2}$ invariants which measure the parity of the number of edge states at each of the particle-hole symmetric momentum and quasi-energy points, $\epsilon T=0,\pi$ and $k=0,\pi$, where $k$ is the dimensionless ribbon momentum.
These can be in general different for a ribbon along the $\textbf{a}_1$ and the $\textbf{a}_2$ directions, although this does not happen for our model \cite{Fulga2019}. 
Since there are two particle-hole symmetric momenta and quasi-energies each, there are in total four weak indices. 
We denote them by $\nu_{k,\epsilon T}$, with $k \in \{ 0, \pi \}$ and $\epsilon T \in \{ 0,\pi \}$.
An odd parity of the number of edge states is indicated by $\nu_{k,\epsilon T}=-1$, whereas an even parity means $\nu_{k,\epsilon T}=+1$.

Note that the above invariants are not all independent. 
They can be related to each other by the relation $\nu_{0,\epsilon T}\nu_{\pi,\epsilon T}=(-1)^{\mathcal{W}_{\epsilon T}}$ \cite{Ran2010}. 
Therefore, there are four independent indices characterizing each phase: $\mathcal{W}_{0}$, $\mathcal{W}_{\pi}$, $\nu_{k=\pi,\epsilon T=0}$, and $\nu_{k=\pi,\epsilon T=\pi}$. 
Also note that, unlike the strong index, the weak invariants require both PHS as well as translation symmetry.
Perturbations which break either of these symmetries will thus convert the weak phases into trivial ones.
Full details of the strong and weak indices for Floquet systems can be found in Ref.~\cite{Fulga2016}.

We describe each of the phases in detail. 
We have labeled them as $\mathfrak{A}$, $\mathfrak{B}$, $\mathfrak{C}$, $\mathfrak{D}$, $\mathfrak{E}$, and $\mathfrak{F}$.
Representative band structures of these phases are shown in Fig.~\ref{fig:phase_diag}.

Phases $\mathfrak{A}$ and $\mathfrak{B}$ are strong phases, which host chiral edge states only in the $\epsilon T=0$ gap, while there are no topologically protected edge modes in the $\epsilon T=\pi$ gap. 
Phase $\mathfrak{B}$ has edge modes with an opposite direction of propagation compared to those phase $\mathfrak{A}$.

Phases $\mathfrak{C}$ and $\mathfrak{D}$ have edge states present at both the $\epsilon T=0$ and $\epsilon T=\pi$ gaps.
The edge states at $\epsilon T=0$ are counter-propagating, whereas those at  $\epsilon T=\pi$ are chiral. 
Therefore, these phases show a combination of strong and weak topology.
Similar to $\mathfrak{A}$ and $\mathfrak{B}$, the chiral edge modes in the $\epsilon T=\pi$ gap of phases $\mathfrak{C}$ and $\mathfrak{D}$ propagate in opposite directions.

Phase $\mathfrak{E}$ is called an anomalous phase, because it hosts topologically protected chiral edges states at both the quasi-energy gaps, even though the Chern number is zero for both of the bulk bands.
Due to this feature, there can be no static counterpart of this phase.
At a resonant hopping strength of $J_{s}T/4=\pi/2$ and $J_d=0$, the bulk bands in the phase become completely flat. 

The phase $\mathfrak{F}$ is a weak phase. 
This phase has trivial strong indices and nontrivial weak indices which are protected by PHS and translation symmetry. 
It has a pair of counter-propagating edge states on each edge at $\epsilon T=0$, and no edge states in the quasi-energy gap at $\epsilon T=\pi$.

By direct inspection of the band structures in each phase, we can enumerate their topological invariants, see Tab.~\ref{tab:invariants}.
Note, however, that all band structures are shown for ribbons with zig-zag edges.
For an armchair edge, the unit cell is two times larger, leading to a folding of the Brillouin zone.
Thus, for an armchair edge, the counter-propagating edge modes overlap in quasi-energy and momentum, and are no longer protected \cite{Rostami2016}. 
This is consistent with the fact that weak topological phases are protected by translation symmetry, since a doubling of the unit cell represents a translation-symmetry-breaking pertubation.

\begin{table}[tb]
\begin{centering}
\begin{tabular}{|c||c|c|c|c|c|c|}
\hline
Phase & $\mathcal{W}_{\epsilon T=0}$ & $\mathcal{W}_{\epsilon T=\pi}$ & $\nu_{k=\pi,\epsilon T=0}$ & $\nu_{k=\pi,\epsilon T=\pi}$ \\
\hline\hline
$\mathfrak{A}$ & $-1$ &  $0$ & $-1$ & $+1$ \\
\hline
$\mathfrak{B}$ & $+1$ &  $0$ & $-1$ & $+1$  \\
\hline
$\mathfrak{C}$ &  $0$ & $+1$ & $-1$ & $+1$ \\
\hline
$\mathfrak{D}$ &  $0$ & $-1$ & $-1$ & $+1$ \\
\hline
$\mathfrak{E}$ & $-1$ & $-1$ & $-1$ & $+1$ \\
\hline 
$\mathfrak{F}$ &  $0$ &  $0$ & $-1$ & $+1$ \\
\hline
\end{tabular}
\caption{
The table of invariants. 
The table lists all the topological invariants of different phases present in the system. 
Each phase is characterized by a set of strong and weak invariants. 
Four invariants uniquely determine each phase.
}
\label{tab:invariants}
\end{centering}
\end{table}

\section{Effects of disorder}\label{sec:disorder}

Topological phases are in general robust to disorders in the system. 
In the presence of disorder, the bulk bands of a two-dimensional system may become localized, provided their Chern numbers vanish. 
However, as the strength of the disorder is increased, the bulk may become delocalized, signaling a transition away from the topological phase. 

We consider a system consisting of $L\times L$ unit cells, indexed by pairs of integers $(n_1, n_2)$, and impose periodic boundary conditions along the ${\bf a}_1$ direction.
We define the scattering matrix at any quasi-energy $\epsilon$ by attaching absorbing terminals to the top and bottom boundaries, $n_2=1$ and $n_2=L$. 
The expression for the scattering matrix is
\begin{equation}\label{eq:smatrix}
S(\epsilon) = P\big[ 1 - e^{i\epsilon}F(1 - P^{T}P )  \big]^{-1} e^{i\epsilon} F P^{T}
\end{equation}
where the superscript $T$ stands for matrix transpose, and we have omitted the time period multiplying $\epsilon$ to avoid confusion. $P$ is a projection operator onto the absorbing terminals,
\begin{equation}
\begin{split}
P = \begin{cases}
1 \quad \text{if} \quad n_{2} \in \{ 1,L \}, \\
0 \quad \text{otherwise.}
\end{cases}
\end{split}
\end{equation}

The above formula for the scattering matrix can be understood by expanding it as a series, 
\begin{equation}
\begin{split}
S(\epsilon) = Pe^{i\epsilon}FP^{T} + Pe^{i\epsilon}FQe^{i\epsilon}FP^{T} + \\
P(e^{i\epsilon}FQ)^{2}e^{i\epsilon}FP^{T} + \ldots
\end{split}
\end{equation}
where $Q=(1 - P^{T}P )$. 
Each of the terms in the above series describes the time evolution over one additional period. 
The states are evolved for one period by applying the Floquet operator ($F$), any portion of the state that overlaps with the absorbing terminal is projected out ($P$), whereas any remaining states ($Q$) are evolved for another period ($F$), etc.
Summing over infinitely many such processes produces the geometric series above, whose total sum is the inverse matrix in Eq.~\eqref{eq:smatrix}.

The scattering matrix has the form 
\begin{equation}
S(\epsilon) = \begin{pmatrix}
\mathfrak{r} & \mathfrak{t} \\
\mathfrak{r}^{\prime} & \mathfrak{t}^{\prime}
\end{pmatrix},
\end{equation} 
where $\mathfrak{t}^{(\prime)}$ and $\mathfrak{r}^{(\prime)}$ are the so-called transmission and reflection blocks. Their elements are the probability amplitudes for states to be transmitted between the two terminals, or backscattered in the same terminal, respectively.
From either of the transmission blocks in the scattering matrix, we can compute the total quasiparticle transmission as
\begin{equation}
G = \text{Tr}(\mathfrak{t}^{\dagger}\mathfrak{t}).
\end{equation}  
In addition, the scattering matrix can be used to compute the strong and weak topological invariants, even in the presence of disorder. We briefly review how this is done in App.~\ref{app:invariants}.

\begin{figure*}[tb]
\includegraphics[width=\textwidth]{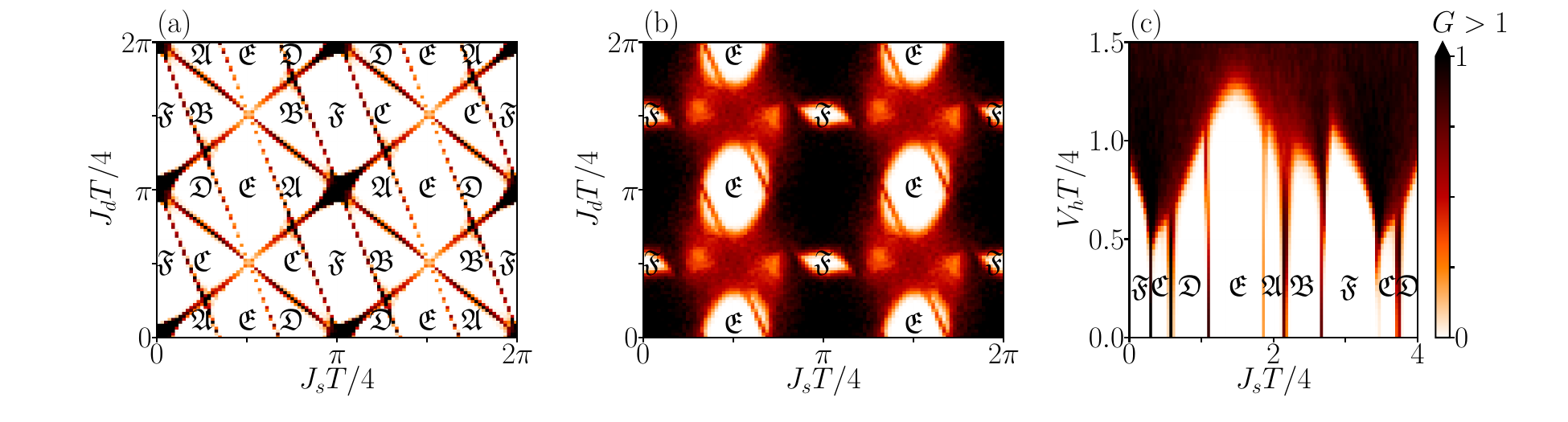}
\caption{
Phase diagram for $V_{h}T/4=0.2$ (a) and $V_{h}T/4=1.0$ (b).
The largest of the transmissions at $\epsilon T=0$ and $\epsilon T=\pi$ is plotted as a function of $J_s$ and $J_d$. 
Each point is obtained by averaging over $100$ disorder configurations. 
(c) Phase diagram for hopping disorder strength $V_{h}T/4$ versus $J_{s}T/4$. 
This is plotted on the path $J_{d}T/4=J_{s}T/4 + 2$. 
Same as in panels (a) and (b), this plot shows larger of the transmissions at $\epsilon T= 0$ and $\epsilon T=\pi$. 
Each point is obtained by averaging over $200$ disorder configurations. 
All the panels have been obtained for a system of size of $24 \times 24$ unit cells with periodic boundary conditions along the $\textbf{a}_1$ direction. 
}
\label{fig:dis.hop}
\end{figure*}

\subsection{Analytic Floquet operator}

From Eq.~\eqref{eq:smatrix} we see that computing the scattering matrix involves writing the Floquet operator in real space. 
To do this, one has to write down the time-evolution operator in each time slice separately and multiply them.
The task of computing the time-evolution operators in general for a Hamiltonian in real space requires numerically exponentiating a large Hermitian matrix, which can be time consuming. 
However, for the specific kind of time drive that we have, one can write down an analytic form of the time-evolution operator in real space for each time slice. 
Hence, we can construct the Floquet operator in real space by not resorting to numerical exponentiation.
This method significantly improves the efficiency of constructing the Floquet operator and hence is useful to do simulations of large systems. 
The details for constructing the analytic Floquet operator are given in Appendix \ref{app:analytic.floquet}. 
In the next two subsections we study the effects of disorder in hoppings and the effects of disorder in the onsite term.
We see how each of the phases behaves for various strengths of hopping as well as onsite disorders. 

\subsection{Hopping Disorder}

We add disorder in the hoppings as 
\begin{equation}\label{eq:dishops}
\begin{split}
J_{s}\frac{T}{4} \rightarrow (J_{s} + \delta_{h})\frac{T}{4}, \\
J_{d}\frac{T}{4} \rightarrow (J_{d} + \delta_{h})\frac{T}{4},
\end{split}
\end{equation}
where $\delta_{h}$ is random number drawn independently for each hopping from the uniform distribution $[-V_{h}, V_{h})$, with $V_h$ the disorder strength. 
Note that hopping disorder breaks all lattice symmetries but preserves PHS, meaning that the weak topological phases remain well defined even as disorder is introduced. This is because, as shown in Refs.~\cite{Ringel2012, Fulga2014, Diez2015, Fulga2016}, weak topological phases can survive the disorder-induced breaking of translation symmetry, provided that local symmetries are preserved exactly, and provided that the ensemble of disordered systems still preserves translation symmetry on average.
Phases protected by average symmetries are dubbed statistical topological phases.

\begin{figure}[tb]
\includegraphics[width=\columnwidth]{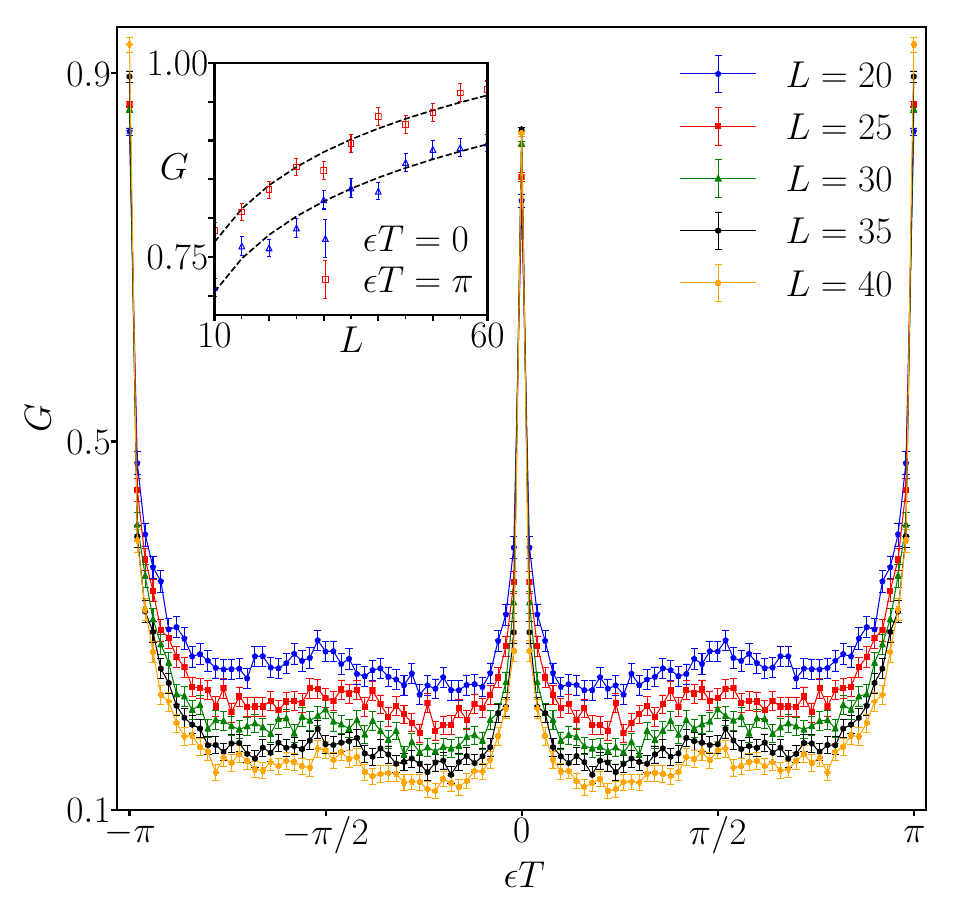}
\caption{
Plot of quasi-energy versus bulk transmission. 
Weak anti-localization can be observed at particle-hole symmetric quasi-energies $\epsilon T=0$ and $\epsilon T=\pi$, where $G$ increases as $\log L$ with system size (inset). 
The transmission values in both the scaling plots and the quasi-energy versus $G$ are obtained by averaging over 1000 disorder configurations. 
All the calculations use $JsT/4=1$, $J_{d}T/4=3$, and $V_{h}T/4=1.5$. }
\label{fig:weak.antiloc}
\end{figure}

In Fig.~\ref{fig:dis.hop} we plot the transmission as a function of hopping strengths $J_{s}$ and $J_{d}$ for two different disorder strengths.
We see that $G$ increases at places where there is a bulk gap closing line in the clean limit. 
This increase in the transmission indicates the existence of delocalized bulk states at the quasi-energies $\epsilon T =0$, $\epsilon T=\pi$, or both.
This shows that a topological phase transition has occurred at those parameter values. 

When the strength of disorder is small, we see that all the phases remain as in the clean limit [see Fig.~\ref{fig:dis.hop}(a)]. 
We have also verified this by calculating the invariants in each phase.
For larger disorder strengths, however, we see that many phases are lost [Fig.~\ref{fig:dis.hop}(b)]. 
Phases $\mathfrak{A}$ and $\mathfrak{B}$ become delocalized at most of the regions in their parameter space except for a small region near to phases $\mathfrak{E}$ and $\mathfrak{F}$. 
A similar property can seen for phases $\mathfrak{C}$ and $\mathfrak{D}$.

\begin{figure*}[tb]
\includegraphics[width=\textwidth]{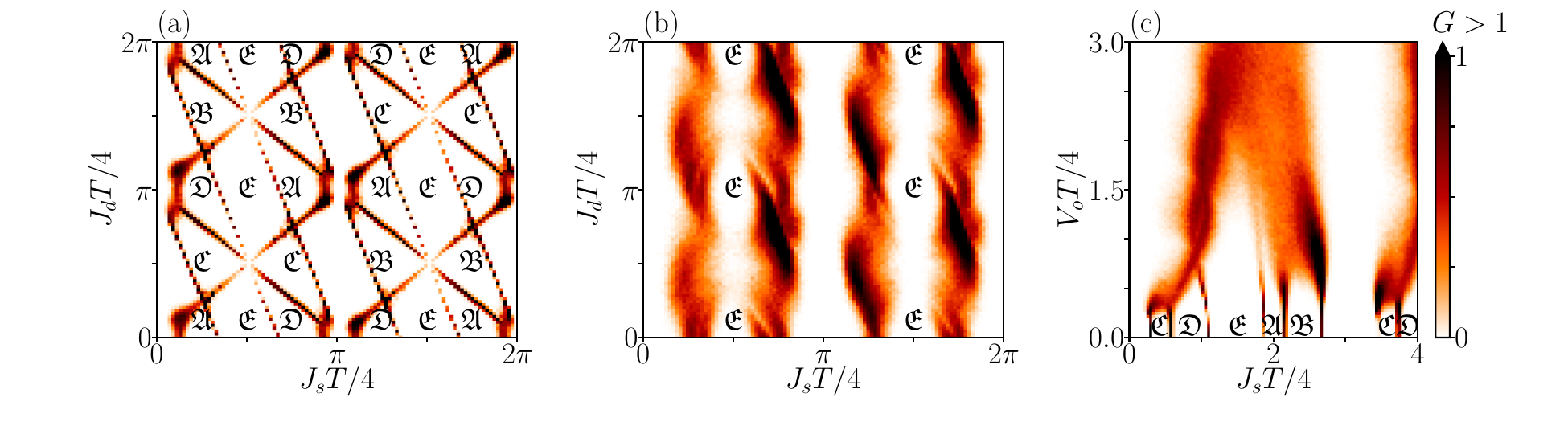}
\caption{
Phase diagram for onsite disorder strength $V_{o}T/4=0.2$ (a) and $V_{o}T/4=1.0$ (b).
The largest of the transmissions at $\epsilon T=0$ and $\epsilon T=\pi$ is plotted as a function of $J_s$ and $J_d$. 
(c) The right most panel shows the transmission along the line $J_{d}T/4=J_{s}T/4+2$ as a function of onsite disorder strength $V_{o}T/4$. 
All the panels have been obtained for a system of size $24 \times 24$ unit cells with periodic boundary conditions along the $\textbf{a}_1$ direction.
In panels (a) and (b), each point has been averaged over 100 disorder configurations, whereas in panel (c) each point has been averaged over 200 disorder configurations.
}
\label{fig:dis.onsite}
\end{figure*} 

Phases $\mathfrak{E}$ and $\mathfrak{F}$, on the other hand, remain localized for some regions of the hopping parameters even for relatively large disorders. 
This is particularly true for phase $\mathfrak{E}$, which still is present in a large region of the $J_s$ -- $J_d$ plane even at disorder strengths for which almost all other phases have disappeared [see Fig.~\ref{fig:dis.hop}(b)]. 
We trace this robustness to the large bulk gap of this phase in the clean limit, close to its so-called ``resonant driving point,'' $\frac{J_{s}T}{4}=\pi/2$ and $\frac{J_{d}T}{4}=0$.
There, the bulk bands are completely flat and positioned at $\epsilon T=\pm\pi/2$, such that the bulk gaps around $\epsilon T=0,\pi$ each occupy one half of the full Floquet quasi-energy zone.

The above analysis shows that large disorders can delocalize the bulk. 
However, this does not give a complete picture of the behavior of phases as a function of disorder strength $V_{h}$. 
To get a better understanding of the robustness of phases to disorder, we choose a path $J_{d}T/4 = J_{s}T/4 + 2$ in the parameter space and plot the transmission at the gap closing quasi-energies $\epsilon T=0$ and $\epsilon T=\pi$ as $J_{s}$ and $V_{h}$ are varied. 
Note that the path chosen covers all the phases as one varies $J_{s}T/4$ from $0$ to $4$. 
Therefore, we can track the behavior of all the phases as a function of the disorder strength. 
A plot of the transmission is shown in Fig.~\ref{fig:dis.hop}(c).
The bulk becomes delocalized above a certain critical disorder strength, which is different for different phases. 
As one can notice, the phases $\mathfrak{E}$ and $\mathfrak{F}$ can withstand a relatively large disorder strengths.

To understand the nature of the phases once the transition has happened, we plot the transmission as a function of quasi-energy and system size in Fig.~\ref{fig:weak.antiloc}. 
We see that at the particle hole symmetric quasi-energies $\epsilon T=0$ and $\epsilon T=\pi$, the transmission increases as a function of system size. 
This peak in the bulk transmission is due to weak anti-localization: a quantum interference effect that enhances bulk transmission \cite{Evers2008}.  
We perform a scaling analysis at these particle hole symmetric quasi-energies, which shows that the system behaves similar to a thermal metal at $\epsilon T= 0$ and $\epsilon T= \pi$, since $G \propto \log(L)$ \cite{Cho1997, Kagalovsky1999, Chalker2001, Evers2008, Medvedyeva2010, Fulga2020} (see inset of Fig.~\ref{fig:weak.antiloc}). 
In contrast, away from the particle-hole symmetric quasi-energies, the transmission is a decreasing function of system size, a consequence of weak localization.

\subsection{Onsite Disorder}

Next we study the effects of onsite disorder, which is added as
\begin{equation}
\nonumber \mu \rightarrow 0 + \delta_{o}\frac{T}{4}.
\end{equation}
The random number $\delta_{o}$ is drawn independently for each site from the uniform distribution of $[-V_{o}, V_{o})$, with $V_o$ the onsite disorder strength. Unlike the hopping disorder considered previously, onsite disorder breaks PHS. 
This means that all weak phases will become trivial as soon as disorder is turned on, and the only remaining topology of the model will be due to its strong topological invariants.

In Fig.~\ref{fig:dis.onsite} we plot the largest of the two bulk transmissions at the quasi-energies $\epsilon T= 0$ and $\epsilon T=\pi$ for various values of $J_{s}$ and $J_{d}$.
Phase $\mathfrak{F}$, which is a weak phase, is lost as soon as disorder is introduced, as discussed above. 

The phases $\mathfrak{A}$, $\mathfrak{B}$, $\mathfrak{C}$, and $\mathfrak{D}$ are robust against small disorder strengths, since they are characterized by nonzero strong invariants. 
However, all of them are eventually destroyed as disorder strength is progressively increased.

Phase $\mathfrak{E}$, the anomalous phase, is robust against a large disorder strength, as can be seen in Fig.~\ref{fig:dis.onsite}. 
As the strength of disorder is increased, the boundaries separating different topological phases move in the $J_s$ -- $J_d$ plane. 
However the middle region of $\mathfrak{E}$ remains localized even for a large disorder, as seen in Fig.~\ref{fig:dis.onsite}(c). 
Since the bulk bands in this phase have a zero Chern number, in phase $\mathfrak{E}$ all bulk states are localized by disorder irrespective of their quasi-energy. 
This is the signature of an AFAI.

\begin{figure}[tb]
\includegraphics[width=\columnwidth]{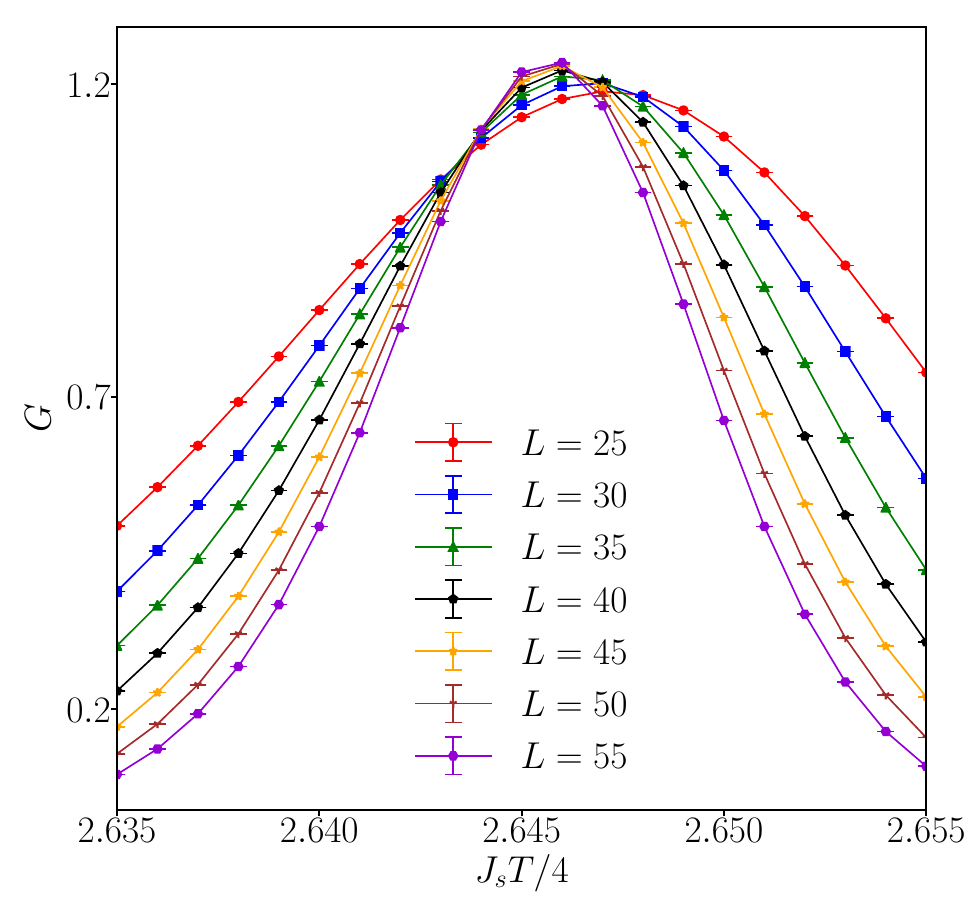}
\caption{Scaling of the bulk transmission across a phase transition, using $J_{d}T/4= J_{s}T/4 + 2$ and $V_{o}T/4=0.2$. 
Each point is obtained by averaging over 1000 disorder configurations, and the lines are guides to the eye.
}
\label{fig:onsite.scaling}
\end{figure}

One prominent feature of Fig.~\ref{fig:dis.onsite} is the presence of large finite-size effects. 
Since onsite disorder breaks PHS, no thermal metal-like phases are possible \cite{Evers2008}, and we expect that the system should always show a localized bulk, with the exception of sharp lines where the strong invariant changes.
Instead, we observe that there are wide regions in parameter space where the finite system shows a significant transmission, and these regions become larger as the strength of disorder is increased.
To show that these regions are indeed finite-size effects, we perform a scaling analysis, as shown in Fig.~\ref{fig:onsite.scaling}.
We find that both the increase in transmission around $J_sT/4=2.645$ and the apparent shift in the maximum transmission occur only for small system sizes.
As the system size is increased, we indeed observe the expected behavior.
The transmission is a decreasing function of $L$ for all values of $J_s$ except at the phase transition point, even though its values remain relatively large for the system sizes we can reach.
This type of quantum-Hall-like phase transition occurs whenever a strong index changes value, so it is present also at the transition between phases $\mathfrak{A}$ and $\mathfrak{B}$, $\mathfrak{E}$ and $\mathfrak{A}$, etc. 

\section{A filter}\label{sec:filter}

Recent experiments in photonic crystals have been able to simulate Floquet systems with a great amount of tunability \cite{Mukherjee2017, Maczewsky2017}.
These experiments hence provide a way to realize many of the phases that we have obtained in previous sections. 
The photonic crystal consists of an array of one-dimensional optical waveguides, through which light can propagate. 
Each waveguide plays the role of a site in the tight-binding model, whereas the distance between waveguides controls the hopping strength. 
The position along the longitudinal direction of the waveguides is the effective time variable of the Floquet system: as light propagates, the distance between adjacent waveguides changes, which emulates a time-periodic hopping strength.

In principle, such photonic crystal platforms allow to freely control both the onsite potential and the hopping of the Floquet system they simulate, by adjusting the radius of each waveguide and the distance between them, respectively. 
However, selectively accessing bulk states and edge states can prove difficult in practice. This is because in conventional experiments, the light injected in one or several waveguides will generally populate a mixture of bulk and edge modes, even if these occur at different quasi-energies. Thus, potential applications of photonic crystals, such as using topological edge states for signal processing \cite{Chong2013}, may be hindered by the spurious contribution of the propagating bulk states.

We show how to overcome these issues by designing a disorder-based filter.
In the previous sections we analyzed the properties of all the phases in the presence of disorder, showing that the anomalous phase, phase $\mathfrak{E}$, remains robust to both hopping disorder and onsite disorder. 
All of the the bulk states are localized for large disorder strengths, even as the edges in a finite sample remain extended. 
This property can be made use of to build filters that allow transmission only of the edges, regardless of whether bulk states are populated or not. 

\begin{figure*}[tb]
\includegraphics[width=\textwidth]{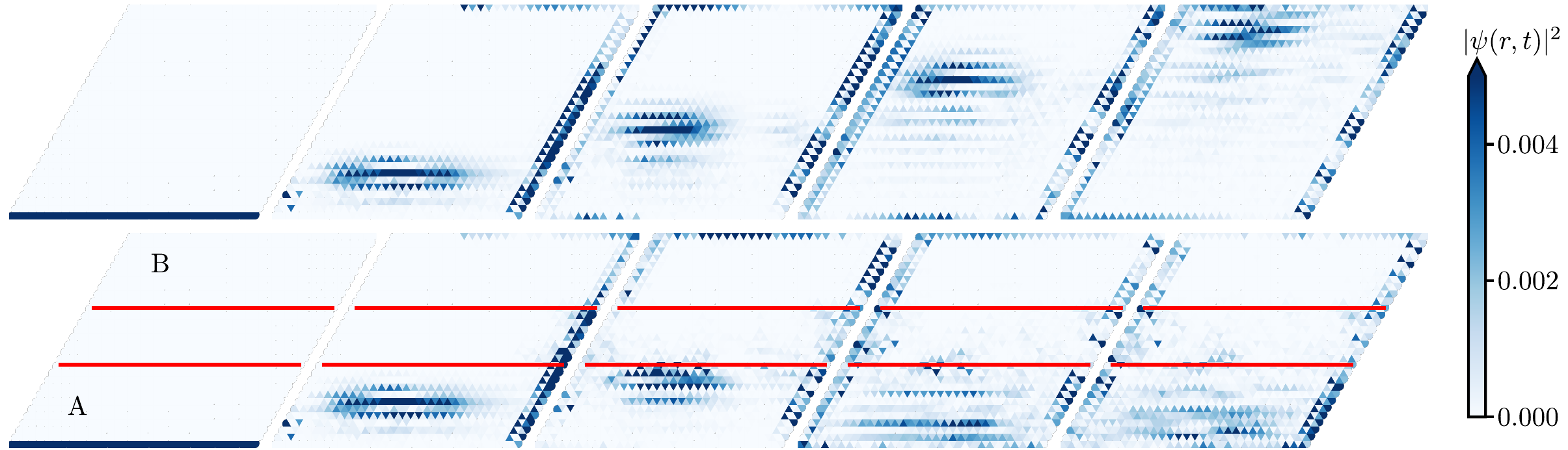}
\caption{
Comparison of the probability density $|\psi|^{2}$ of the evolving state after $t=0$, $50T$, $100T$, $150T$, and $200T$ (from left to right) with and without filter for a system of size $L\times W=30\times 30$.
The top panels show the system without filter. 
The bottom panels show the system with filter. 
The bottom left most panel shows the regions A and B. 
The region in between the red lines in the bottom panels shows the disordered region which acts as a filter. 
The supplemental material contains two movies of this time evolution \cite{code}. 
}
\label{fig:filter.snapshot}
\end{figure*}

We show the action of the filter in a system of $30 \times 30$ unit cells in phase $\mathfrak{E}$ ($J_{s}T/4=1.5$, $J_{d}T/4= 3$). Disorder is present in the central region of the system, in the interval $n_2 \in [12,20]$, where $n_2$ labels the coordinate of the unit cells along the ${\bf a}_2$ direction. 
Hence, the disordered region divides the entire system into three regions: the central, filter region, the bottom region, labeled A, in the interval $n_2 \in [1,11]$, and the top region, labeled B, in the interval $n_2\in [21,30]$ (see  Fig.~\ref{fig:filter.snapshot}).

To simulate the initialization procedure commonly used in photonic crystals, we choose a wavefunction in the region A such that it is spread entirely on the bottom most edge. 
This initial wavefunction is time-evolved by acting on it with the full Floquet operator. 
It begins propagating through both the bulk as well as the edge of the system, since it is a mixture of both bulk and edge modes.
However, as it reaches the disordered region, all of the bulk states are stopped and only the edge states continue to propagate.

The disordered region filters out the signals carried by the bulk states and allows signals to pass only along the edges. 
We quantify the effectiveness with which the bulk states are filtered out by defining the filter coefficient $\eta_{t}$ after an amount of time $t$ as 
\begin{equation}
\eta_{t} = \sum_{r \in \text{bulk of region B}} |\psi(r)|^{2}.
\end{equation}
Here, the bulk of region B denotes all the sites in this region, except for the last two unit cells along the boundary of the sample in region B. 

\begin{figure}[tb]
\includegraphics[width=\columnwidth]{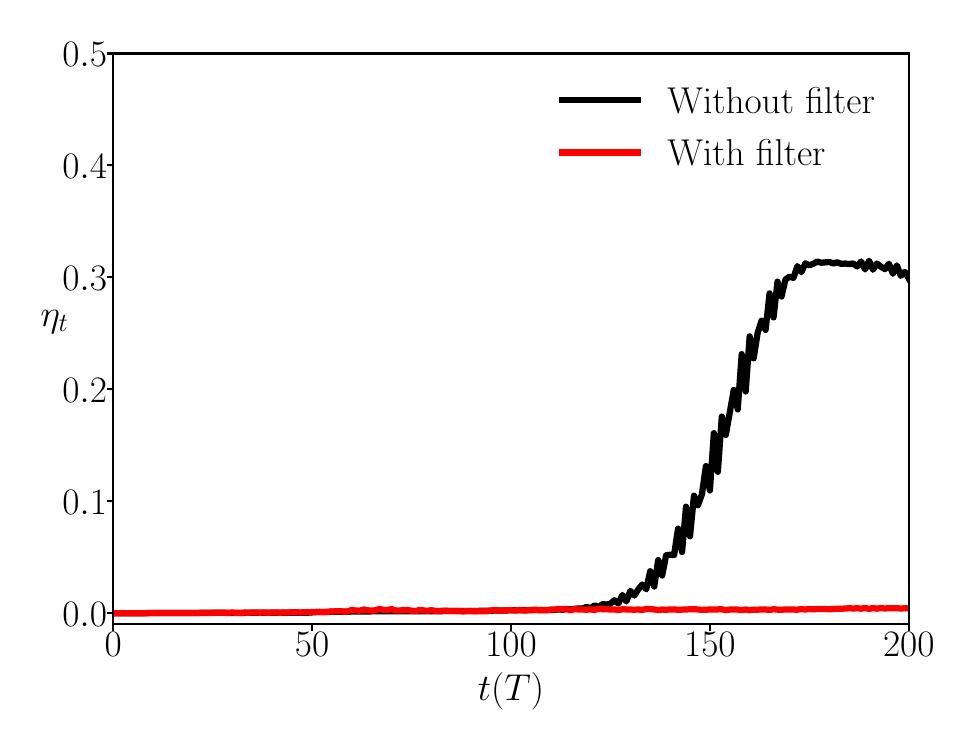}
\caption{
Filter coefficient $\eta_{t}$ as a function of $t$ in the anomalous phase with parameters $\left( J_{s}T/4,J_{d}T/4 \right)\equiv\left(1.5, 3 \right)$. 
The red curve shows the coefficient when the filter is present. In this case the disorder strengths are $\left(V_{h}T/4, V_{o}T/4 \right)=\left(0.2, 1 \right)$. 
The black curve shows the coefficient when the filter is not present.}
\label{fig:filter.coeff}
\end{figure}

When we have an ideal filter, $\eta_{t} = 0$, because the filter allows only the edges to transmit. 
For the simulation that we have carried out with $V_{h}T/4=0.2$ and $V_{o}T/4=1.0$, we find that $\eta_{t=200T} \approx 0.004$.
On the other hand, when no filter is present, the probability of finding the initial state in the bulk of region B increases, leading to $\eta_{t=200T}\approx 0.30$ (see Fig.~\ref{fig:filter.coeff}). Thus, even for a relatively narrow disordered region, the filter has an efficiency of roughly two orders of magnitude.

Note that one of the preconditions for an efficient filter is that the system realizes an anomalous topological phase, since this guarantees both chiral edge modes as well as a disorder-induced localization of all bulk states. 
If the system would instead host a Floquet Chern insulating phase, such as phase $\mathfrak{A}$, then some of the bulk modes would necessarily propagate through the disordered region. 
This is because bands with nonzero Chern numbers cannot be fully localized by disorder: there must exist at least one delocalized bulk state which ``carries the Chern number.''

\section{Conclusions}
\label{sec:conc}

In this work, we have considered a topological phase belonging to the symmetry class BDI. 
The time drive we consider breaks the time-reversal symmetry in the system and hence only particle-hole symmetry is preserved.
In the clean limit, this system possesses in total of six different Floquet topological phases. 
Each of these phases is characterized by four independent topological invariants, two strong indices and two weak indices.
Among these six phases there is an anomalous phase as well as a weak phase that has trivial strong indices. 

After the classification of the clean system we introduced disorder. 
First, we studied the hopping disorder which preserves the particle-hole symmetry. 
An in-depth analysis shows that all the phases are robust to these disorders, i.e.~the invariants remain robust. It was shown that the anomalous phase can withstand more disorder than other phases. At large hopping disorder, weak anti-localization leads to a thermal metal-like phase. 

Next, onsite disorder was studied. 
The onsite disorder breaks the particle-hole symmetry in the problem. 
Hence, the weak phase becomes trivial. 
However, all the other phases were shown to be robust to onsite disorder. 
Again, the anomalous phase was shown to be the most robust phase. 

The anomalous phase is the most robust phase, a fact that can be used in making filters. 
In this anomalous Floquet Anderson insulating phase, the presence of disorder fully localizes the bulk states, with edge states remaining extended along the edges. 
We made use of this property and simulated a filter that allowed signals to pass only through the edges.
We also calculated its filter coefficient to quantify the effectiveness of the filter. 
In the future, it would be interesting to study whether the enhanced robustness of the anomalous phase is a property shared not just by strong and weak Floquet topological phases, but also for Floquet topological crystalline phases \cite{Ladovrechis2019} and Floquet higher-order topological phases \cite{Zhu2021}.

Disorder in real systems is ubiquitous, resulting in unequal hopping strength between neighboring sites on a lattice. 
Earlier experiments \cite{Rechtsman2013, Maczewsky2017, Mukherjee2017} with optical waveguide arrays have opened up the possibility to mimic such Floquet systems. 
One way of manipulating the hopping would be to vary the distance between different waveguide arrays. 
The onsite potential can be changed by varying the radius of the waveguides. 
Motivated by these experimental platforms, our proposal for a disordered based filter is feasible to realize in photonic systems.

\acknowledgments
The authors would like to thank Ulrike Nitzsche for technical support. This work was supported by the Deutsche Forschungsgemeinschaft~(DFG, German
Research Foundation) under Germany's Excellence Strategy through the
W\"{u}rzburg-Dresden Cluster of Excellence on Complexity and Topology in Quantum Matter -- \emph{ct.qmat} (EXC 2147, project-id 390858490), as well as through the DFG grant FU 1253/1-1.

\bibliography{ref}

\begin{appendix}

\section{Scattering matrix invariants}
\label{app:invariants}

We determine the strong and weak topological invariants from the two-terminal scattering matrix defined in the main text.
To this end, we apply twisted periodic boundary conditions to the $L\times W$ unit cell system along the ${\bf a}_1$ direction, setting:
\begin{equation}
\ket{1, n_2} = e^{i\phi} \ket{L, n_2},
\end{equation}
where, as before, $\ket{n_1, n_2}$ labels a state positioned on the unit cell with coordinates $(n_1, n_2)$, $n_{1,2}\in\mathbb{Z}$.
The phase $\phi$ determines the kind of boundary conditions present.  
For $\phi=0$ we have the periodic boundary conditions and for $\phi=\pi$ we have anti-periodic boundary conditions. 

Since the Floquet operator is now a function of $\phi$, the scattering matrix and its reflection block $\mathfrak{r}$ also depend on this twist angle.
This enables us to determine the strong topological index as \cite{Fulga2016}:
\begin{equation}
\mathcal{W}_{\epsilon} = \frac{1}{2\pi i} \int_{0}^{2\pi} d\phi~~\frac{d}{d\phi} \log \det \mathfrak{r}(\phi, \epsilon).
\end{equation}

The weak indices, on the other hand, are given by the sign of the determinant. 
These can be computed by

\begin{equation}
\begin{split}
\nu_{k=0, \epsilon} & = \text{sign} \det \mathfrak{r}(\phi=0, \epsilon),  \\
\nu_{k=\pi, \epsilon} & = \text{sign} \det \mathfrak{r}(\phi=\pi, \epsilon). \\
\end{split}
\end{equation}

\section{Analytic Floquet operator}\label{app:analytic.floquet}

\begin{figure}
\includegraphics[width=0.5\columnwidth]{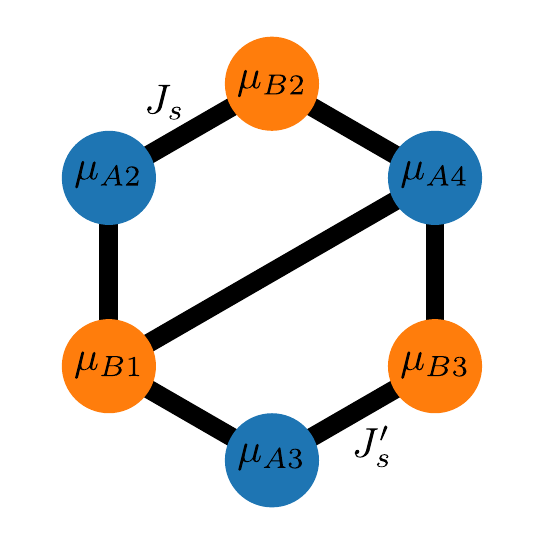}
\caption{A single plaquette, showing the nonzero onsite terms and hopping terms. In the time interval $0 < t < T/4$, only the hoppings marked $J_s$ and $J_s'$ are active.}
\label{fig:syst_app}
\end{figure}

The time-evolution operators in each driving step of the Floquet operator Eq.~\eqref{eq:fq.def} can be determined analytically, even in the most general case in which both the hoppings and the onsite potential are random.
For the purpose of illustration, we show how this is done in the case of a single plaquette (Fig.~\ref{fig:syst_app}). 
The Hamiltonian in the time interval $0 < t < T/4$ for random complex hoppings and a random onsite terms is given by
\begin{equation}\label{eq:basis.ham}
H_{1} = \begin{pmatrix}
\mu_{A2} & 0 & 0 & 0 & J_{s} & 0 \\
0 & \mu_{A3} & 0 & 0 & 0 & J^{\prime}_{s} \\
0 & 0 & \mu_{A4} & 0 & 0 & 0 \\
0 & 0 & 0 & \mu_{B1} & 0 & 0 \\
J^{*}_{s} & 0 & 0 & 0 & \mu_{B2} & 0 \\
0 & J^{\prime *}_{s} & 0 & 0 & 0 & \mu_{B3} 
\end{pmatrix}.
\end{equation}

With an appropriate basis transformation, we can write the above Hamiltonian in a block diagonal form as
\begin{equation}
\tilde{H}_{1} = \begin{pmatrix}
\mu_{B1} & 0 & 0 & 0 & 0 & 0 \\
0 & \mu_{A2} & J_{s} & 0 & 0 & 0 \\
0 & J_{s}^{*} & \mu_{B2} & 0 & 0 & 0 \\
0 & 0 & 0 & \mu_{A3} & J^{\prime}_{s} & 0 \\
0 & 0 & 0 & J_{s}^{\prime *} & \mu_{B3} & 0 \\
0 & 0 & 0 & 0 & 0 & \mu_{A4} 
\end{pmatrix}.
\end{equation}
Each of the $2 \times 2$ blocks have the form
\begin{equation}
h = \begin{pmatrix}
\mu_{A} & J \\
J^{*} & \mu_{B}
\end{pmatrix},
\end{equation}
where $J$ is the hopping strength and $\mu_{A}$ and $\mu_{B}$ are the onsite terms.
Since the exponential of a block diagonal matrix is a block diagonal matrix consisting of the exponentials of the individual blocks, it is enough to see what happens to the exponential of single block:
\begin{equation}
\begin{split}
& \exp \left[ -i\begin{pmatrix}
\mu_{A} & J \\
J^{*} & \mu_{B} 
\end{pmatrix}\frac{T}{4} \right] 
= e^{-i\frac{\mu_{A} + \mu_{B}}{2}\times\frac{T}{4}} ~~~~~ \times \\
& \begin{pmatrix}
\cos(|\textbf{p}|)-i\frac{\mu}{|\textbf{p}|}\sin(|\textbf{p}|) & -i\frac{|JT/4|}{|\textbf{p}|}\sin(|\textbf{p}|)e^{i\phi} \\
-i\frac{|JT/4|}{|\textbf{p}|}\sin(|\textbf{p}|)e^{-i\phi} & \cos(|\textbf{p}|)+ i\frac{\mu}{|\textbf{p}|}\sin(|\textbf{p}|)
\end{pmatrix},
\end{split}
\end{equation}
where $\mu = (\mu_{A} - \mu_{B}) T/8$, $\phi = \arg(JT/4)$ and $\textbf{p} = (|JT/4|\cos(\phi), -|JT/4|\sin(\phi), \mu/2 )$. 
The time-evolution operator $\tilde{F}_{1}$ is obtained by replacing the blocks with the above exponentiated blocks.
However, since this time-evolution operator is not in the original basis, one can do a reverse basis transformation to obtain $F_1$ in the original basis of Eq.~\eqref{eq:basis.ham}. 

Note that this matrix can be constructed explicitly for different hopping strengths.  
A matrix such as the one above is constructed for each time slice and multiplied to obtain the complete Floquet operator. 
Since this form of constructing the Floquet operator does not require numerical exponentiation, this method can speed up the process of finding the Floquet operator.
Also, this is an efficient method when disorder is present.
The strengths of hopping $J_{s}$ and the onsite terms $\mu_{A}, \mu_{B}$ can be independently varied. 
Hence, the Floquet matrix can be constructed efficiently. 

\end{appendix}

\end{document}